\documentclass[twocolumn,preprintnumbers,endnote,nofootinbib,aps,prd,eqsecnum,floatfix]{revtex4}

\usepackage{color}

\usepackage{footmisc,multirow}
\usepackage{subfigure}
\usepackage{amsmath,amssymb,slashed}

\usepackage{graphicx}

\newcommand{\be}{\begin{eqnarray*}}
\newcommand{\ee}{\end{eqnarray*}}
\newcommand{\gl}[1]{(\ref{#1})}

\newcommand{\bee}{\begin{eqnarray}}
\newcommand{\eee}{\end{eqnarray}}
\newcommand{\beeq}{\begin{equation}}
\newcommand{\eeeq}{\end{equation}}
\newcommand{\gev}{{\rm{GeV}}}
\newcommand{\tev}{~{\rm{TeV}}}

\newcommand{\citeb}[1]{\hbox{\cite{#1}}}

\begin{document}

\title{New physics in LHC Higgs boson pair production}

\begin{abstract}
  Multi-Higgs production provides a phenomenologically clear window to
  the electroweak symmetry breaking sector. We perform a comprehensive
  and comparative analysis of new electroweak physics effects in
  di-Higgs and di-Higgs+jet production. In particular, we discuss
  resonant di-Higgs phenomenology, which arises in the Higgs portal
  model and in the MSSM at small $\tan\beta$, and non-resonant new
  physics contributions to di-Higgs production in models where the
  newly discovered Higgs candidate is interpreted as a
  pseudo-Nambu-Goldstone boson. We show that, for all these scenarios,
  a measurement of the di-Higgs and di-Higgs+jet final states provides
  an accessible and elaborate handle to understand electroweak
  symmetry breaking in great detail.
\end{abstract}

\author{Matthew J. Dolan\vspace{0.2cm}} \email{m.j.dolan@durham.ac.uk}
\affiliation{Institute for Particle Physics Phenomenology, Department
  of Physics,\\Durham University, Durham DH1 3LE, United Kingdom}
\author{Christoph Englert} \email{christoph.englert@durham.ac.uk}
\affiliation{Institute for Particle Physics Phenomenology, Department
  of Physics,\\Durham University, Durham DH1 3LE, United Kingdom}
\author{Michael Spannowsky} \email{michael.spannowsky@durham.ac.uk}
\affiliation{Institute for Particle Physics Phenomenology, Department
  of Physics,\\Durham University, Durham DH1 3LE, United Kingdom}
\pacs{}

\preprint{IPPP/12/80}
\preprint{DCPT/12/160}

\maketitle

\section{Introduction}
\label{sec:intro}
Both ATLAS and CMS have observed a Standard Model-like Higgs
boson~\citeb{Higgs} at around $125~\gev$~\citeb{atlas,cms}. In the
very same mass region, the combination of the D$\slashed{0}$ and CDF
collaborations' data sets exhibits a SM-like Higgs excess with a local
significance of 2.2$\sigma$~\citeb{tevat}. The implications of this
newly-discovered particle have already been discussed in the context
of the SM and beyond~\citeb{fits,plehnino,bsm}. The combined local
significance is mostly driven by an excess in the diphoton invariant
mass, consistent with the SM Higgs boson within $2\sigma$. Therefore,
we can expect that the observed particle bears some resemblance to the
SM Higgs since $gg \to h \to \gamma\gamma$ is sensitive to the special
role of the Higgs particle in both the SM's gauge and Yukawa sectors
and their interplay. Correlating this observation with electroweak
precision data~\citeb{LEP2} and measurements in the $h\to ZZ,W^+W^-$
channels, which constrain the particle's couplings to massive gauge
bosons, we infer from fits to the data~\citeb{fits,plehnino} (most
notably by the ATLAS themselves~\citeb{recatlas}) that the particle
reproduces SM Higgs properties within 1-2$\sigma$. This agreement
partially relies on biasing the fit towards the SM Higgs hypothesis by
assuming a total decay width $\Gamma(h\to\hbox{anything})
\simeq\Gamma_h^{\text{SM}}$~\citeb{recatlas} and the absence of new
degrees of freedom in $gg,\gamma\gamma\to h$. These assumptions are,
strictly speaking, neither theoretically nor experimentally motivated.
A precise determination of the particle's couplings relaxing such
assumptions is an LHC lifetime achievement, which will combine direct
searches for heavy states that potentially run in production and decay
loops and constraints of non-standard Higgs branching fractions.

Deviations from the SM Higgs phenomenology even at the 10\% level
leave a lot of space for modifications of the Higgs sector by
Beyond-the-SM (BSM) physics: new physics of roughly that size is
largely unconstrained by the precise investigations of the SM at the
$Z$ mass pole. Given that the corresponding BSM couplings need to be
small, the current data does not provide constraints on weakly-coupled
Higgs sector extensions beyond what we have already learned from
LEP~\citeb{LEP2}. Currently, Monte Carlo-based analyses which target
non-standard decays of the Higgs-like
resonances~\citeb{Englert:2011us,pinv, gluons1,hadrons,taus} suggest
that branching ratio limits of $\lesssim{\cal{O}}(10\%)$ can in
principle be obtained at the LHC from direct measurements, depending
on the characteristics of the non-standard decay. This bound might be
too loose to efficiently probe interactions beyond the SM.

From this perspective, it is imperative to directly probe potential
modifications of the electroweak symmetry breaking sector, if
phenomenologically possible, to fully exhaust the LHC's search
potential to physics beyond the SM. One class of hadron collider
processes which precisely serve this purpose is multi-Higgs
production~\citeb{Plehn:1996wb}. These processes are functions of the
symmetry breaking potential's parameters and are, consequently, highly
sensitive to the realization of electroweak symmetry breaking. While
triple Higgs production is beyond the reach of the LHC
experiments~\citeb{Plehn:2005nk}, di-Higgs production can potentially
be measured in rare decays $pp\to hh\to b\bar b
\gamma\gamma$~\citeb{Baur:2003gp}.  Only recently, the application of
jet substructure techniques~\citeb{Butterworth:2008iy} to di-Higgs
production in boosted final states has uncovered sensitivity in $pp\to
hh(j) \to b\bar b \tau^+\tau^-(+j)$ to both di-Higgs production and
the trilinear Higgs coupling~\citeb{Dolan:2012rv}. This approach is
currently also being investigated by ATLAS \cite{atlashh} in the context of
a LHC luminosity upgrade.

Crucial to the findings of Ref.~\citeb{Dolan:2012rv} is accessing the
small invariant di-Higgs mass phase space region which is mostly
sensitive to the Higgs trilinear coupling with moderately boosted
Higgses $p_T\sim 100~\gev$. The sensitivity can be augmented by
accessing collinear di-Higgs configurations by recoiling the di-Higgs
system against a hard jet~\citeb{Dolan:2012rv}. This configuration is
extremely sensitive to modifications of the trilinear Higgs coupling
since it does not suffer from the kinematical shortcomings that are
present in the inclusive di-Higgs final state, where the Higgs
particles are produced back-to-back. Promising results to measure the
di-Higgs cross section have also been found for extremely boosted
$b\bar b W^+W^-$ production~\citeb{Papaefstathiou:2012qe}.

Motivated by the recently-unravelled sensitivity to di-Higgs
production at the LHC, we perform a comprehensive and comparative
analysis of new physics interactions in LHC di-Higgs and di-Higgs+jet
production in this paper. We divide our discussion into two parts. We
discuss resonant di-Higgs(+jet) signatures in Sec.~\ref{sec:portal},
where we first analyze a simple extension of the Higgs sector via the
so-called Higgs portal~\citeb{wells}. We subsequently discuss
prospects to constrain the MSSM Higgs sector at low $\tan\beta$ via
resonant production of a heavy Higgs $H$ decaying to $hh$.

In Sec.~\ref{sec:pseudo} we discuss the phenomenology of non-resonant
new physics contributions to di-Higgs production in composite Higgs
and dilaton models (to make this work self-contained we briefly
introduce the basics before we comment on the phenomenology). This
broad class of pseudo-Nambu-Goldstone theories comprises many
interesting features in a phenomenologically well-defined framework.
Both these models introduce new degrees of freedom to di-Higgs and
di-Higgs+jet production and modified trilinear couplings compared to
the SM, while the composite Higgs scenarios also introduce new
$t\bar{t}hh$ interactions.  Comparing these models to the SM
expectation provides a consistent framework to constrain the
electroweak symmetry breaking potential with future measurements at
the $\sqrt{s}=14~\tev$ LHC.

Throughout this paper, we produce events and leading order cross
sections using an in-house Monte Carlo code that is based on the
{\sc{Vbfnlo}}~\citeb{vbfnlo} and
{\sc{FeynArts}}/{\sc{FormCalc}}/{\sc{LoopTools}}~\citeb{hahn}
frameworks.

\section{Resonant new physics: From the Higgs Portal to Supersymmetry}

\subsection{Di-Higgs production in the Higgs portal scenario}
\label{sec:portal}
The Higgs portal scenario~\citeb{wells} is a convenient and
theoretically consistent way to generalize the SM in its mostly
unconstrained parameters (such as Higgs boson's total and hidden decay
width) in a minimal approach~\citeb{portal2}. Realizing that
$\Phi^\dagger_S \Phi_S$ transforms as a gauge singlet, where $\Phi_S$
is the SM Higgs doublet, there is a plethora of SM extensions with highly
modified and interesting LHC
phenomenology~\citeb{Englert:2011us,pinv,frank}.  In a 'mirrored'
approach~\citeb{Barbieri:2005ri} the Higgs portal potential reads
\begin{multline}
  \label{eq:potential}
  V = \mu^2_S |\Phi_S|^2 + \lambda_S |\Phi_S|^4 +
  \mu^2_H |\Phi_H|^2 + \lambda_H |\Phi_H|^4\\
   +  \eta_\chi |\Phi_S|^2 |\Phi_H|^2 \,,
\end{multline}
where we have introduced a hidden sector Higgs field $\Phi_H$. The
Higgs portal model allows to identify a viable dark matter candidate
in the hidden sector~\citeb{Kanemura:2010sh}, whose potential LHC
phenomenology has been explored in~\citeb{darkimplications}.

\begin{figure}[t!]
  \includegraphics[width=0.33\textwidth]{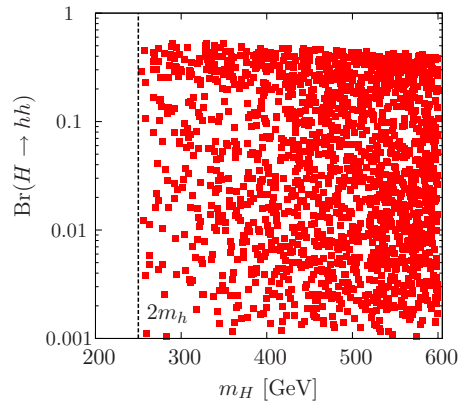}
  \caption{\label{fig:branching} Mass of the heavy Higgs state $H$ if
    $m_{h}=125~\gev$, and consistency with
    $S,T,U$~\citeb{Peskin:1991sw}, unitarity and current ATLAS/CMS
    results is imposed. The density of the model points must not be
    interpreted as a probability measure.}
\end{figure}

%
\begin{figure*}[thp!]
  \subfigure[][~$\sigma/\sigma(\text{portal})$ and invariant di-Higgs
  mass distribution for $pp\to hh+X$ at the LHC $14$ TeV.]{
    \parbox{0.3\textwidth}{
      \includegraphics[width=0.29\textwidth]{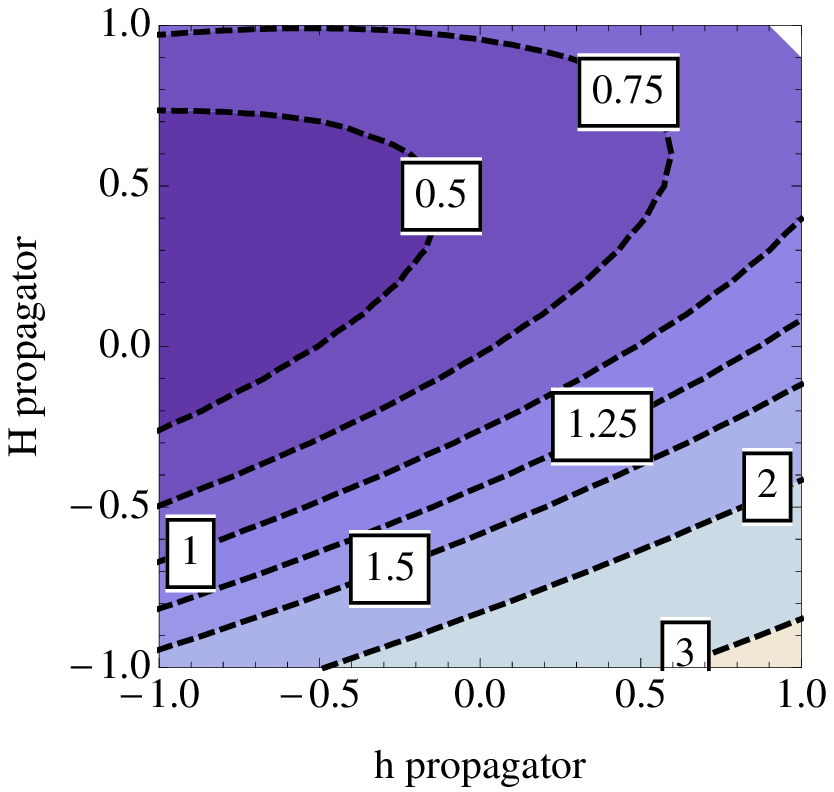}\\[0.2cm]
      \includegraphics[width=0.30\textwidth]{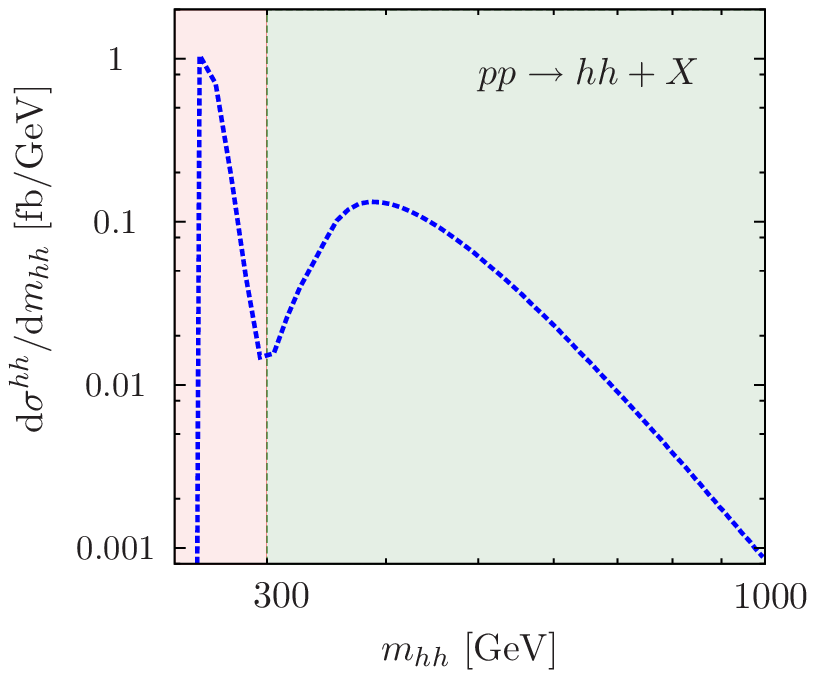}\\[0.2cm]
    }}\hfill \subfigure[][~$\sigma/\sigma(\text{portal})$ and
  invariant di-Higgs mass distribution for $pp\to hH+X$ at the LHC
  $14$ TeV.]{
    \parbox{0.3\textwidth}{
      \includegraphics[width=0.29\textwidth]{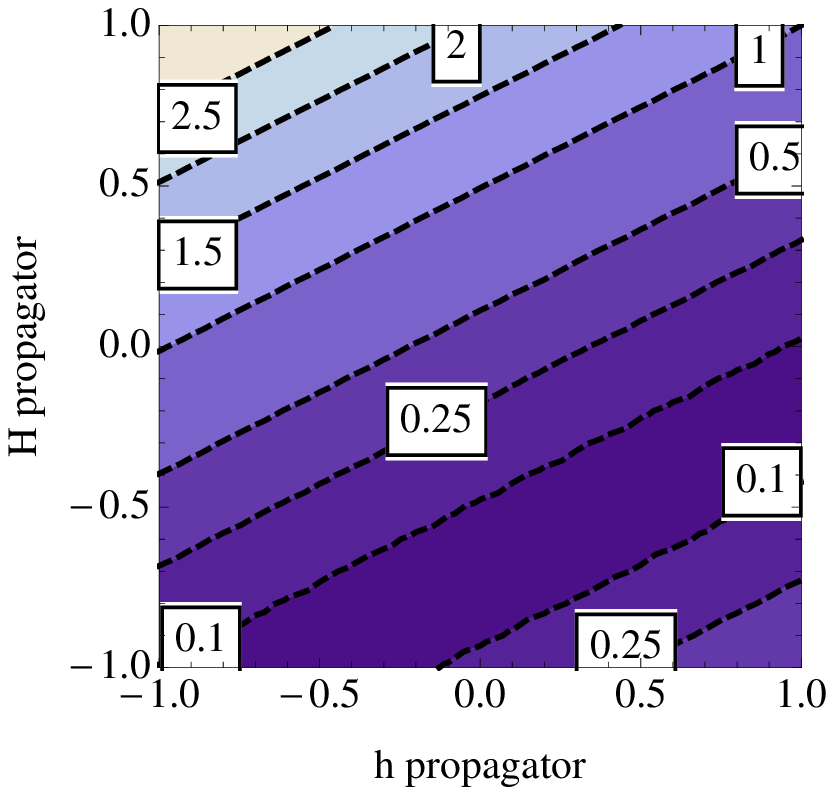}\\[0.2cm]
      \includegraphics[width=0.30\textwidth]{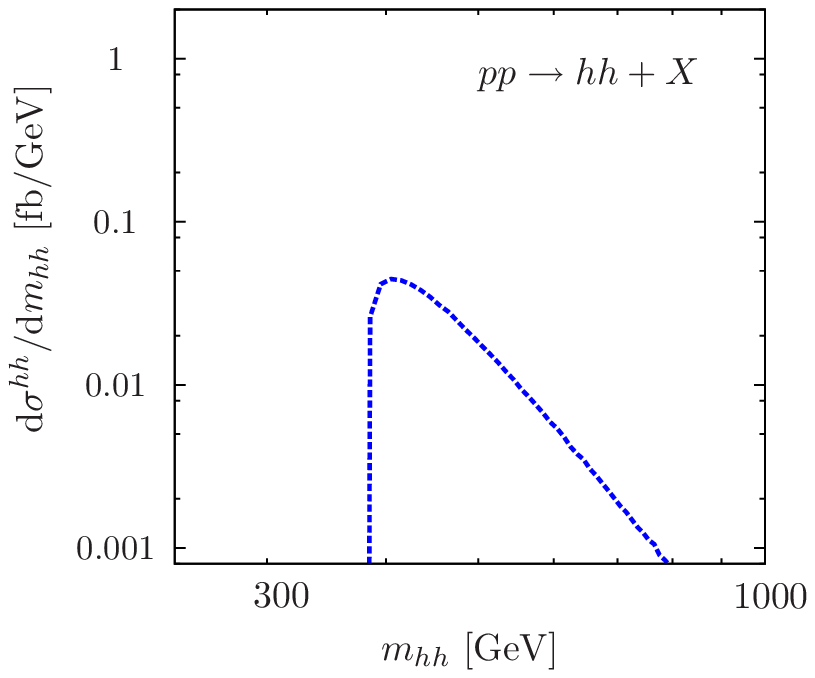}\\[0.2cm]
    }}\hfill \subfigure[][~$\sigma/\sigma(\text{portal})$ and
  invariant di-Higgs mass distribution for $pp\to HH+X$ at the LHC
  $14$ TeV.]{
    \parbox{0.3\textwidth}{          
      \includegraphics[width=0.29\textwidth]{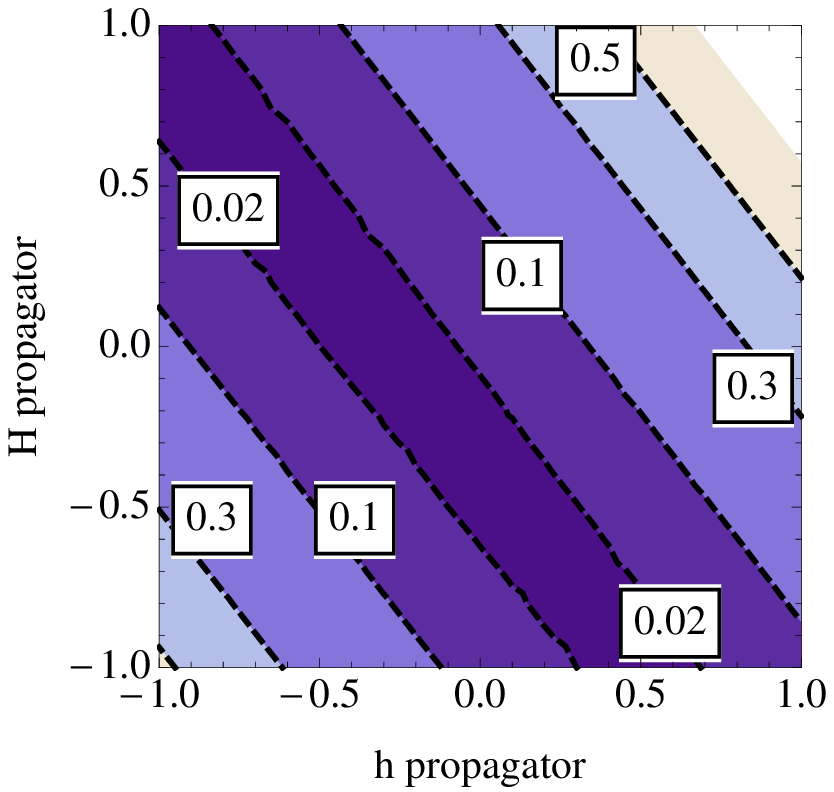}\\[0.2cm]
      \includegraphics[width=0.30\textwidth]{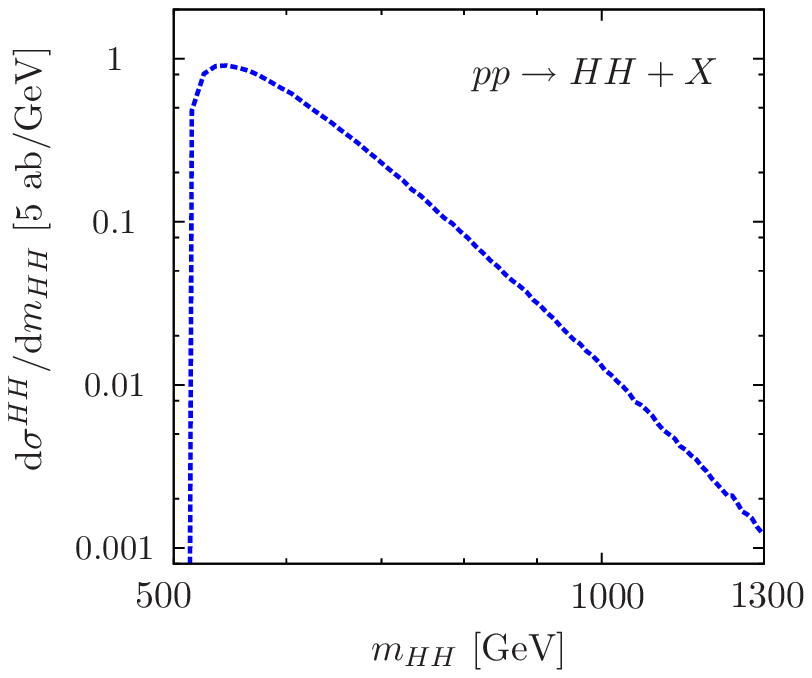}\\[0.2cm]
    }}
  \caption{\label{fig:portalxescs} Upper panels: cross sections in the
    portal scenario for the parameter point mentioned in the text.  We
    scan over the multiples of the trilinear couplings
    Eq.~\gl{eq:tricouplings} that are in one-to-one correspondence
    with diagrams involving the $h,H$ propagators and show contours
    relative to the central expectation Eq.~\gl{eq:portalres}. Lower
    panels: invariant di-Higgs mass distributions.}
\end{figure*}

After symmetry breaking, which is triggered by the Higgs fields
acquiring vacuum expectation values (vevs)
$|\Phi_{S,H}|=v_{S,H}/\sqrt{2}$, the would-be-Nambu-Goldstone bosons
are eaten by the $W^\pm,Z$ fields and the corresponding directions in
the hidden gauge sector, and the only effect (in unitary gauge) is a
two-dimensional isometry which mixes the visible and the hidden Higgs
bosons:
\begin{equation}
  \begin{split}
    h =& \phantom{-} \cos\chi \, H_s + \sin\chi \, H_h     \\
    H =& -\sin\chi \, H_s + \cos\chi \, H_h  \,,
    \label{eq:mixi}
  \end{split}
\end{equation}
where $\chi$ is a function of the portal potential parameters
Eq.~\gl{eq:potential} (for details see, {\it
  e.g.},~\citeb{portal2}). For the remainder of this section we choose
$m_H>m_h=125~\gev$.

Electroweak precision measurements and unitarity requirements of
longitudinal gauge boson scattering and massive quark annihilation to
longitudinal gauge bosons suggest that, if such a model is realized in
nature, then the mixing should preferably be far from maximal,
$\cos\chi^2\approx 1$, which for generic {\it{perturbative}} choices
of the potential $\lambda_{S},\lambda_{V},\eta_\chi\ll 4\pi$ results
in a typically small mass splitting between the physical Higgs states
$h,H$. Admitting some tuning, a larger mass splitting can be arranged,
which results in a clean LHC phenomenology of narrow trans-TeV
resonances~\citeb{wells2}.  Small mass splittings imply a
phenomenologically more involved situation since the light Higgs
bosons are produced with small transverse momentum in di-Higgs
production. Nonetheless, given the vastly enriched Higgs sector
phenomenology, we can still study the Higgs portal in sufficient
detail to fully reconstruct the Higgs potential
Eq.~\gl{eq:potential}~\citeb{portal2}. Crucial in this reconstruction
algorithm is the measurement of the invisible Higgs decay branching
ratio~\citeb{dieter1,Englert:2011us}. It can be immensely improved by a
possible observation of a cascade decay $H\to hh$. Additional
information from observing all multi-Higgs signatures (and the
trilinear couplings especially), if phenomenologically accessible, can
be used to further constrain or even rule out the simple portal
extension.

Expanding Eq.~\gl{eq:potential} around the vacuum expectation values,
we get the trilinear couplings relevant for di-Higgs
production\footnote{Triple Higgs production, which is sensitive to the
  modified Higgs quartic couplings yields phenomenologically
  irrelevant cross sections just like in the SM~\citeb{Plehn:2005nk}.}:
\begin{subequations}
  \label{eq:tricouplings}
  \begin{align}
    hhh: \quad & {3/2}( \nonumber 2 \lambda_H s_\chi^3 v_H + 2
    \lambda_S c_\chi^3 v_S\\ & \hspace{1.5cm}
    +\eta_\chi c_\chi^2 s_\chi v_H +\eta_\chi  c_\chi s_\chi^2 v_S)\,, \\
    HHH: \quad & {3/2} (2 \lambda_H c_\chi^3 v_H - 2 \lambda_S
    s_\chi^3 v_S \nonumber \\ & \hspace{1.5cm}
    + \eta_\chi c_\chi  s_\chi^2 v_H - \eta_\chi c_\chi^2  s_\chi v_S) \,,\\
    hHH: \quad & 2 ( 3\lambda_H-\eta_\chi ) c_\chi^2 s_\chi v_h +
    2( 3\lambda_S-\eta_\chi) c_\chi s_\chi^2v_S \nonumber \\ &
    \hspace{2.cm}
    +\eta_\chi  s_\chi^3v_H + \eta_\chi c_\chi^3  v_S \,,\\
    hhH: \quad & 2  ( 3 \lambda_H- \eta_\chi) c_\chi  s_\chi^2 v_H - 2 ( 3
    \lambda_S- \eta_\chi ) c_\chi^2  s_\chi v_S  \nonumber \\ &
    \hspace{2.cm} 
    +  \eta_\chi c_\chi^3 v_H - \eta_\chi s_\chi^3 v_S\,,
  \end{align}
\end{subequations}
where $c_{\chi}=\cos\chi$ and $s_{\chi}=\sin\chi$. Current
observations leave open a lot of parameter space for such signatures
to be relevant at the LHC. In Fig.~\ref{fig:branching} we scan over
the parameters of the Higgs portal potential enforcing unitarity and
electroweak precision constraints, as well as current limits from the
ATLAS and CMS experiments~\citeb{atlas,cms}. If the heavier Higgs mass
is $m_H\geq 250~\gev$, there are parameter choices such that the
$\sin^2\chi$ suppression of the $H$ decay to SM matter from
Eq.~\gl{eq:mixi} renders the prompt decay of $H$ to observable SM
matter subdominant to the cascade decay $H\to hh$. This can be traced
back to large trilinear couplings ${\cal{O}}(v_H,v_S)$ that arise as a
consequence of electroweak symmetry breaking. Therefore, there is the
possibility to constrain the portal model by measuring the trilinear
couplings in resonant and non-resonant $pp\to hh,hH,HH+X$ production.

\begin{figure}[t!]
  \includegraphics[width=0.33\textwidth]{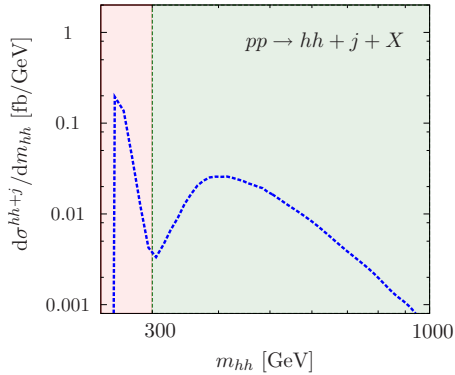}
  \caption{\label{fig:smallj} Invariant mass distribution for $pp\to
    hh+j+X$ in the portal scenario.}
\end{figure}

In Fig.~\ref{fig:portalxescs}, we show a scan over the cross sections
of $pp\to hh,hH,HH\to~(\text{visible})$ as functions of the involved
trilinear couplings for a exemplary parameter point $v_S\simeq
246~\gev$, $v_H\simeq 24~\gev$, $m_h=125~\gev$, and $m_H\simeq
255~\gev$, $\Gamma_H=24~\gev$. The central inclusive cross section
values at leading order implying (prompt) visible final states are
\begin{subequations}
  \label{eq:portalres}
  \begin{align}
    pp\to hh+X \quad  : & \quad  44.4~{\text{fb}} \\
    pp\to Hh+X \quad : & \quad 5.57~{\text{fb}} \\
    pp\to HH+X \quad :  & \quad 667~{\text{ab}}
  \end{align}
\end{subequations}
 (the SM cross section is 16~fb). Comparing to the NLO QCD corrections
in the context of the (MS)SM by running {\sc{Higlu}}~\citeb{higlu} and
{\sc{Hpair}}~\citeb{hpair}, we can expect an enhancement of the cross
section by about $K=\sigma^{\text{NLO}}/\sigma^{\text{LO}}\simeq
2$. 

For $pp\to hh+j+X$ with $p_{T,j}\geq 80~\gev$ we calculate a
leading-order cross section of $\sigma=10.1~\text{fb}$
(Fig.~\ref{fig:smallj}) which should be contrasted to a SM
leading-order cross section of $\sigma=2.58~\text{fb}$, which can be
isolated from the background~\citeb{Dolan:2012rv}. Hence, the
measurement of the one jet-inclusive cross section will assist in
formulating constraints on such a model.

Note that, $pp\to HH+X\to {\text{visible}}$ is naively suppressed
$\sim\sin^6\chi$. Therefore, for the bulk of the portal parameter
space, heavy di-Higgs production (and di-Higgs+jet production
different from $pp\to hh+j+X$) is phenomenologically inaccessible at
too small rates, with no space left for kinematical
signal-over-background $S/B$ improvements.

\bigskip

{\it Summary:} The Higgs portal scenario offers the possibility of
large enhancements in the di-Higgs production rate, from both resonant
and non-resonant (via changes in $\lambda_{hhh}$) new
physics. Extracting the rate for $pp\to h^* \to hh$ using the boosted
kinematical techniques from our previous paper~\cite{Dolan:2012rv}
along with measuring the resonant peak in the di-Higgs invariant mass
spectrum will aid in the full reconstruction of the Higgs portal
lagrangian by correlating these two independent measurements. This
strategy is assisted by the cross section's large dependence on
$\lambda_{hhh}$. A high luminosity analysis of $hh$ and $hh+j$
production can also facilitate a measurement of the trilinear coupling
in this model.

\subsection{The MSSM at small $\tan\beta$}
\label{sec:mssm}
The Higgs portal model of Sec.~\ref{sec:portal} bears some resemblance
to a generic two Higgs doublet model, and therefore our findings are
also relevant for searches for supersymmetry in the context of the
MSSM and its extensions. 

The trilinear couplings of the Higgs bosons in the MSSM are given by
\begin{equation}
  \begin{split}
\lambda_{hhh} =&  \phantom{-} 3 \cos 2\alpha \, \sin (\beta + \alpha)  \\
\lambda_{Hhh} =&  \phantom{-}  2\sin 2\alpha \, \sin (\beta + \alpha) - \cos 2 \alpha \, \cos(\beta + \alpha)  \\
\lambda_{HHh} =& - 2\sin 2 \alpha \, \cos (\beta + \alpha) - \cos 2 \alpha \, \sin (\beta + \alpha),
  \end{split}
\end{equation}
up to radiative corrections, details of which can be found in the
second reference of~\cite{Plehn:1996wb}, $\tan\beta=v_u / v_d$ is the
ratio of vevs of the two MSSM Higgs doublets, and $\alpha$
diagonalizes the Higgs mixing matrix. The above couplings are in units
of $\lambda_0 = M_Z^2/v$. In principle, disentangling the
contributions proportional to $\lambda_{Hhh}$ and $\lambda_{hhh}$ in
double Higgs production would allow a reconstruction of the angles
$\alpha$ and $\beta$.

We observe that when $\beta$ is small and we are near the decoupling
limit where $\alpha \sim \beta - \pi/2$ then the $\lambda_{Hhh}$ is
proportional to $\cos\beta$. Thus when $2 m_h < m_H < 2 m_t$ $H$ has a
large branching ratio into a pair of Higgses $hh$, similar to the
Higgs portal model in Sec.~\ref{sec:portal}. 
Probing the dihiggs final states is thus probably the best way of finding $H$ if $\tan\beta$
is low.  The presence of squarks can further enhanced
the production by running in the loops\footnote{Note that, depending
  on the color charge assignment di-Higgs production can be enhanced
  compared to single Higgs production~\citeb{Kribs:2012kz}.}.

Achieving a Higgs mass of 125~GeV at such low values of $\tan\beta$
requires exceptionally heavy stop masses and mixings. Scanning over
the squark masses and mixings, we find that $m_{\tilde{q}}>50$~TeV in
order to achieve $m_h \sim 125$~GeV. These spectra are characteristic
of 'mini-split' SUSY, which has recently been advocated
in~\cite{minisplit}, which suggests that the weak scale is tuned and
supersymmetry present at higher energies.  However, it is unusual to
have all the scalars heavy except the extra Higgses. This would
require the presence of a cancellation between $m_{H_u}^2$ and
$m_{H_d}^2$ if these quantities are large like the other scalar soft
terms, or else that they have some suppression relative to the squark
and slepton masses.

\begin{figure}[t!]
  \subfigure[][]{
    \parbox{0.33\textwidth}{
      \includegraphics[width=0.33\textwidth]{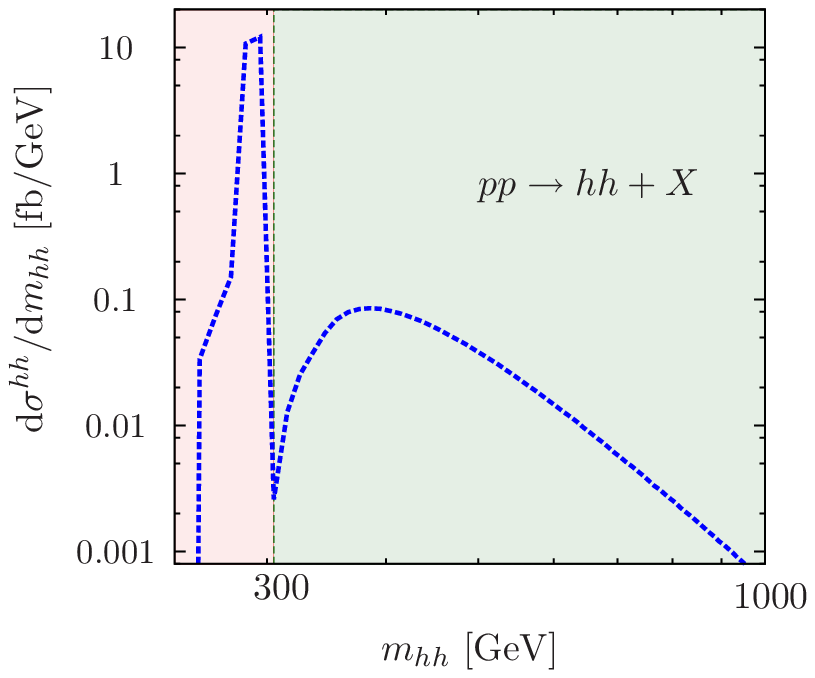}\\[0.3cm]
      }
  }
  \vspace{0.2cm}
  \subfigure[][]{
    \parbox{0.33\textwidth}{
    \includegraphics[width=0.33\textwidth]{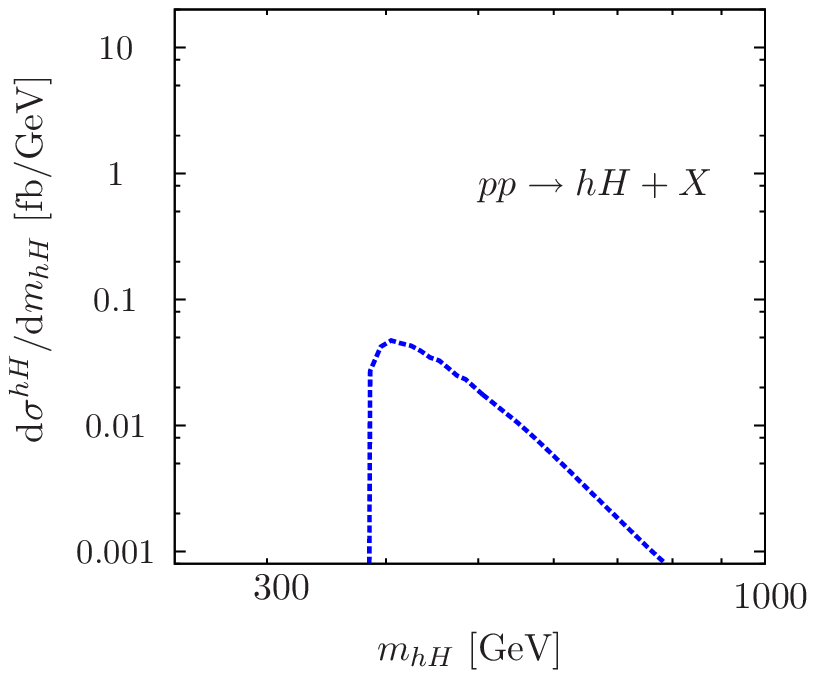}\\[0.3cm]
    }
    }
  \caption{\label{fig:susy_branching} Invariant mass distribution of the
    (a) $hh$ and the (b) $Hh$ system for MSSM-like production at low
    $tan\beta$. For details see text.}
\end{figure}

Moving beyond the MSSM, another possibility is that $\tan\beta$ is low
and that a large contribution to the mass of the lightest (SM-like)
Higgs boson comes from an extra singlet field $S$ with superpotential
couplings $\lambda S H_u H_d$. This induces an extra contribution to
the Higgs mass $\propto \lambda^2 \sin^2 2\beta$ which is enhanced at
low $\tan\beta$; this is the so-called $\lambda$-SUSY scenario of the
NMSSM~\cite{lambdasusy}. We focus exclusively on the MSSM here,
however we expect the phenomenology to be similar in the NMSSM if the
singlet-like states are heavier than the MSSM Higgses.

We find a point with $\tan\beta =3 $, and adjust the scalar masses
until we achieve $m_h \sim 125$~GeV. We set the mass of the other
CP-even boson of the MSSM $H$ to be 290~GeV. In this regime the
branching fraction ${\text{BR}}(H \to hh) \sim 45\% $, and the decay
width is $\Gamma_H =0.25$~GeV. The other main partial decays are into
$b\bar{b}$ (12\%), $W^+ W^-$ (28\%) and $ZZ$ (12\%).  We could
increase the branching ratio into two Higgses further by decreasing
$\tan\beta$, at the cost of increasing the scalar masses. Using a
suitably modified version of \textsc{Vbfnlo} we find the leading order
production cross-section $\sigma(pp\to H \to hh)= 246$~fb.  We also
calculate the cross-section for $\sigma ( pp \to H \to Hh)$. This is
suppressed by the off-shell $H$ in the s-channel, and by the fact that
the $\lambda_{HHh}$ coupling is suppressed relative to the
$\lambda_{Hhh}$ coupling. We find the cross-section for this process
to be 4.5~fb, too low for observation given $h$ has SM-Higgs-like
branching ratios.

We can separate the large contribution $H \to hh$ by reconstructing
the di-Higgs invariant mass which exhibits a peak at $m_H$. This
allows us to extract the cross-section for $pp\to H \to hh$, and after
cutting around the peak the remainder of the events are due to $pp\to
h \to hh$. As in the Higgs portal model, this process can be extracted
using the techniques from our previous paper, allowing constraints to
be put on $\alpha$ and $\beta$. The invariant mass distribution and
rate for the $hh+j$ final state are also similar to the portal
scenario, Fig.~\ref{fig:smallj}

\bigskip

{\it Summary:} The di-Higgs phenomenology in the MSSM at low
$\tan\beta$ is similar in many respects to that of the Higgs portal
model. Measurements of the resonant and non-resonant contributions to
di-Higgs production allows a reconstruction of the parameters $\alpha$
and $\beta$.

\section{Nonresonant new physics: Pseudo-Nambu-Goldstoneism}
\label{sec:pseudo}
Apart from softly-broken supersymmetry, strong interactions are the
only other constructions which can cure the naturalness problem (if
only partially) with phenomenologically testable implications. 

A well-known example of electroweak symmetry breaking from strong
interactions is technicolor (TC) where $m_W\sim f$ where $f$ is the
``pion'' decay constant. The techni-$\Sigma$ and techni-$\rho$
resonances will have masses of the order of the TC confining scale,
which can be much larger than the electroweak scale,
$\Lambda_{\text{TC}}\gg f$. This usually triggers a tension with
curing the quadratic energy divergence in perturbative longitudinal
gauge boson scattering, which demands at least a single light degree
of freedom. An illustrative example which incorporates such a state is
easily constructed from the holographic interpretation of a bulk gauge
theory broken by boundary conditions in a Randall-Sundrum background
\citeb{Csaki:2003zu}\footnote{Owing to the large $N$ and large 't
  Hooft coupling limit \cite{hooft} of AdS/CFT, it is intrinsically
  difficult to construct a fully realistic model in terms of
  electroweak precision measurements.}: The appearance of the infrared
brane signals the spontaneous breakdown of conformal invariance in the
dual picture \cite{Rattazzi:2000hs}. This is accompanied by higgsing
of a symmetry, which is weakly gauged into the strongly-interacting
sector. On the one hand, such a ``higgsless'' theory does not have
light scalar degrees of freedom analogous to the SM Higgs boson. On
the other hand, stabilizing the compactification moduli via the
Goldberger-Wise mechanism \citeb{Randall:1999ee} lifts the zero mass
radion, which couples to the conformal anomaly
\begin{multline}
  \label{eq:coupling}
  T^\mu_\mu\sim m_W^2W^+_\mu W^{-\,\mu}+{m_w^2\over \cos^2\theta_w}Z_\mu
  Z^\mu \\ + \sum_f m_f \bar f f + \dots \,.
\end{multline}
In the CFT picture we identify a pseudo-dilaton, which has an
impressive resemblance to the SM Higgs boson as a consequence of its
couplings. In this sense, the dilaton mimics a light
Higgs boson because the mass terms are the source of scaling
violation.

Different to this approach is the interpretation of the entire Higgs
multiplet as a set of Nambu-Goldstone bosons. There are multiple ways
to construct such a model consistently, ranging from collective
symmetry breaking~\citeb{perelstein} to holographic Higgs
models~\citeb{Contino:2003ve,Giudice:2007fh} which vary in their
details and symmetry content. Common to all these realizations is the
breaking of a global symmetry pattern by gauging a subgroup of the
strongly interacting sector.

While there are parameter choices for both scenarios which are
consistent with the SM in their {\it{single}} Higgs phenomenology, the
measurement of the di-Higgs(+jet) production can be a key
discriminator between these different non-resonant realizations.

\subsection{Di-dilaton production}
\label{sec:dilaton}

We first discuss the implications of interpreting the 125 GeV boson as
a pseudo-dilaton~\citeb{chacko,more-dilaton}.  We note that there is a
substantial number of options in modelling the electroweak sector
using strong interactions, and thus the conclusions of this section
should be taken as illustrative rather than definitive for this class
of models.

The pseudo-dilaton is associated with the spontaneous breaking of
scale symmetry at an unknown scale $f$, and we denote this field by
$\chi$. The couplings of the pseudo-dilaton to massive Standard Model
particles are determined by its coupling to the trace of the SM energy
momentum-tensor $T_{\mu\nu}$, Eq.~\gl{eq:coupling}. The couplings of
the dilaton to the massive SM particles are thus the same as those of
the SM Higgs, but rescaled by a factor of $v/f$. The couplings of the
pseudo-dilaton to gluons and photons are given by
\begin{equation}
  {\cal{L}}^{D5}_{\chi,\text{massless}}=\frac{\alpha_{EM}}{8 \pi f}c_{EM} 
  \chi F_{\mu\nu}F^{\mu\nu} + \frac{\alpha_{S}}{8\pi f} c_S 
  \chi G_{\mu\nu}^{a}G^{a \mu\nu}
  \label{eq:massless_dil_coupling}
\end{equation}
where $c_{EM,G}$ are anomaly coefficients. The precise value these
take depends on what further assumptions are made about the UV
dynamics of the theory and what heavy colored and electromagnetically
charged states are present. We assume that the dilaton couples to
photons and gluons via the full QCD/EM
beta-function~\citeb{Goldberger:2007zk,Fan:2008jk}. We also consider
the model of~\citeb{Campbell:2011iw} which studies the same system
with an extra family of quarks. In both these cases we can find
SM-like behavior with an enhanced $\sigma \times {\text{BR}}$ into
photon pairs.

In the fully composite scenario, the dilaton couples to massless gauge
bosons via the full beta-function, we have $c_{EM}=-17/9$ and $c_S =
11- 2n_f /3$ where $n_f=5$~\citeb{Goldberger:2007zk} is the number of
light quarks. In the four-family model we have $c_{EM}=-6/5$ and
$c_s=4/3$, and obtain a similar single-dilaton phenomenology. The
production cross-section of the dilaton can be enhanced by orders of
magnitude relative to the SM value. However, as the dominant decay
channel then becomes $\chi \to gg$, the cross-section times branching
ratio of the observable final states $\chi \to f\bar{f}$, $\chi \to
VV$ and $\chi \to \gamma \gamma$ can still be close to their SM
values, depending on the scale $f$ (following arguments similar to the
ones presented in
Ref.~\citeb{Englert:2011us,Coleppa:2011zx,Barger:2011nu}).

There will also generally be dimension six operators,
the most interesting of which is~\citeb{Manohar:2006gz}
\begin{equation}
  {\cal{L}}^{D6} = -\frac{\alpha_s}{4\pi f^2} c_{\chi\chi GG} \chi^2 (G_{\mu\nu}^a)^2\,.
  \label{eq:dilaton_D6}
\end{equation}
We define the $D6$ operator with a minus sign, so that $c_{\chi\chi
  GG} >0 $ complies with the low-energy effective Higgs
theorems~\citeb{Shifman:1979eb,higlu,kniehl} paradigm: Integrating out
the heavy top quark, we obtain an effective interaction ${\cal{L}}\sim
G^a_{\mu\nu}G^{a\, \mu\nu} \log(1+h/v)$ in the SM.

It is important to keep in mind that the higher dimensional
interactions with the gluon and the photon fields arise from
integrating out the conformal dynamics and need not follow the LET
paradigm, which predicts a unique coupling structure of the $h^n
G^{a\,2}_{\mu\nu}$ interactions as a consequence of $m\propto
\left\langle h \right \rangle$ for all fundamental masses in the
SM\footnote{It is intriguing to realize that there is, in fact, a
  connection between LET and the vanishing trace anomaly
  Eq.~\gl{eq:coupling} for infrared photons $\lim_{Q^2\to
    0}\left\langle 0| T^\mu_\mu | \gamma\gamma\right\rangle = 0$
  \cite{Adler:1976zt,kniehl}.}. 
We also explore the possibility that the dimension
six operator is negligible, by setting $c_{\chi\chi GG }=0$.

\begin{figure}[!t]
  \includegraphics[width=0.35\textwidth]{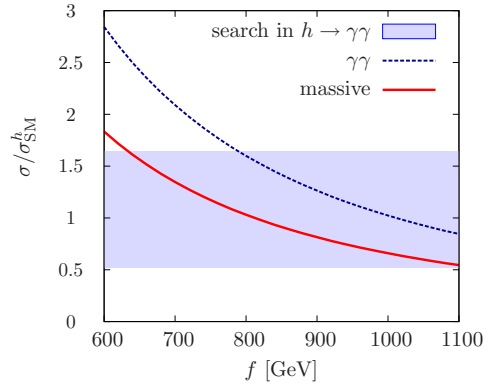}
  \caption{\label{fig:dilprod} Dilaton production from gluon fusion
    with current limits of the $h\gamma \gamma$ coupling analysis
    \cite{plehnino} included.}
\end{figure}

For the fully composite model we find that for $f=850$~GeV that
$\sigma \times {\text{BR}}$ of the dilaton into massive final states
are very similar to those in the Standard Model, and $\chi \to \gamma
\gamma$ it is 1.55 times the Standard Model value. For gluons we find
$\sigma \times {\text{BR}}$ is enhanced by a factor of approximately
150.  This agrees with values obtained from recent fits of
experimental data in~\citeb{fits,plehnino,chacko}. We show in
Fig.~\ref{fig:dilprod} the $\sigma \times {\text{BR}}$ for massive
states and for $\gamma\gamma$, normalized to the SM values . We also
include a blue horizontal band indicating the signal strength in the
diphoton channels from combining the ATLAS and CMS
searches~\cite{atlas,cms}. In the four-family case we obtain similar
results for $f \sim 570$~GeV.

The dilaton's total decay width is approximately 5~MeV and very
similar to the SM. Upper limits on the Higgs width are difficult to
assess experimentally \cite{Dobrescu:2012td} and will eventually be
limited by large systematic uncertainties
\cite{DeRoeck:2009id}. Constraints on the dilaton model arising from
such measurements will be too loose to rule this model out. As we will
see, investigating multi-dilaton production provides the missing
handle to constrain the model when consistency with single Higgs
observations prevails.

\begin{figure*}[thp!]
  \begin{center}
    \parbox{0.4\textwidth}{
      \includegraphics[width=0.4\textwidth]{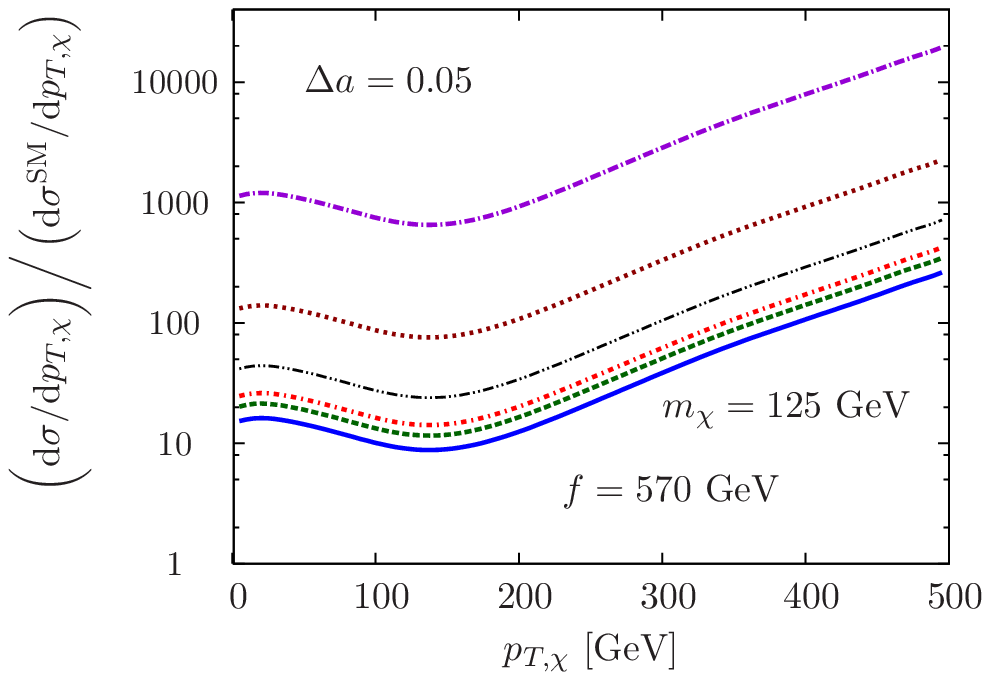}
    }
    \hspace{1.2cm}
    \parbox{0.4\textwidth}{
      \includegraphics[width=0.4\textwidth]{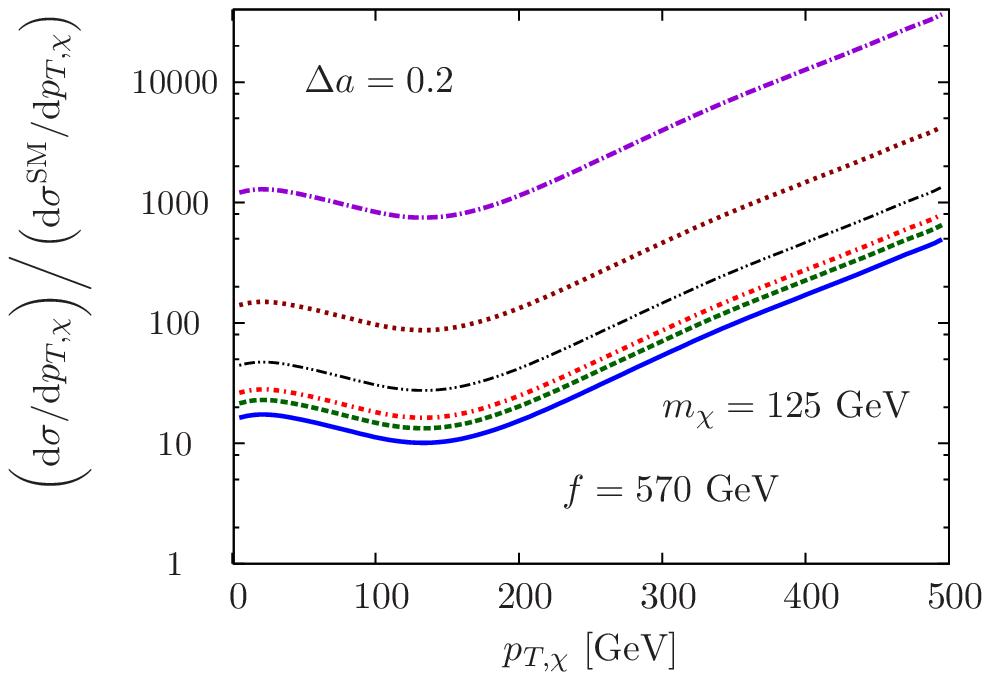} 
    }
    \parbox{0.4\textwidth}{
      \vspace{0.25cm}
      \includegraphics[width=0.4\textwidth]{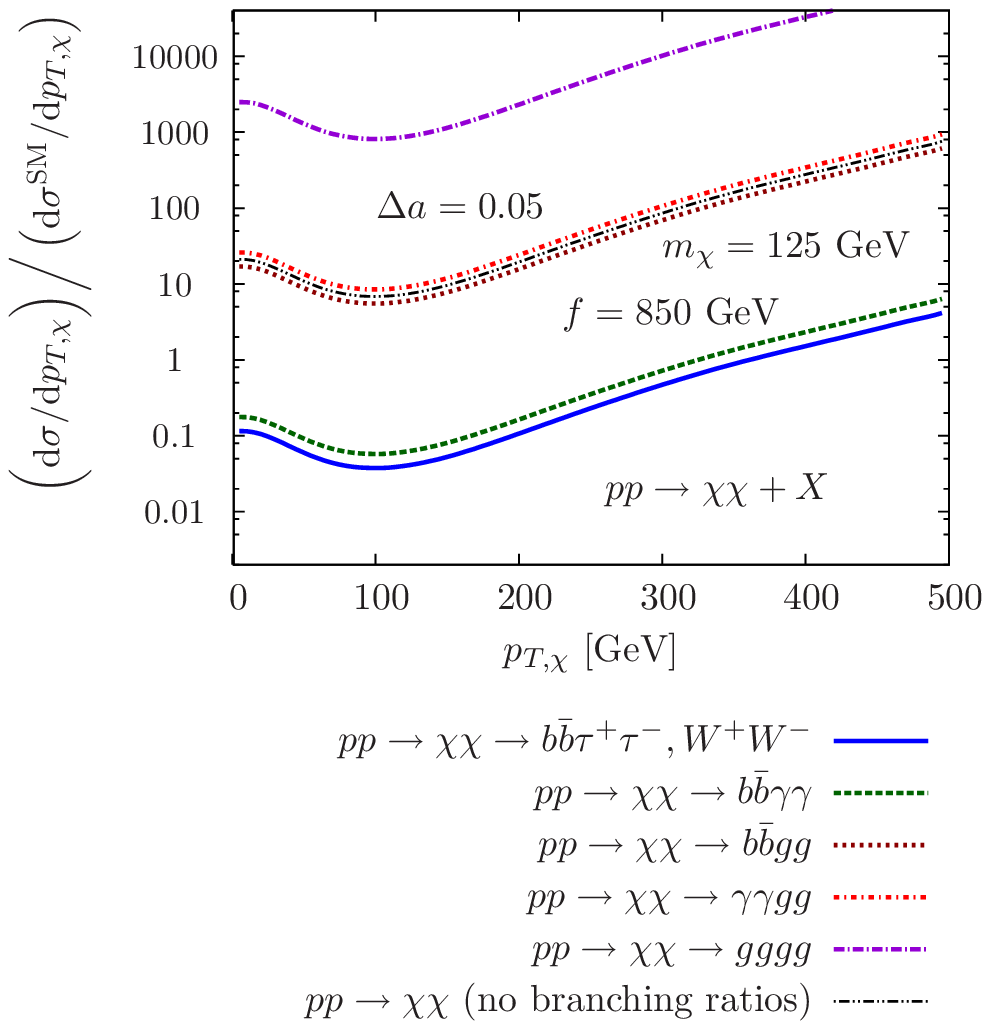}
    }
    \hspace{1.2cm}
    \parbox{0.4\textwidth}{
      \includegraphics[width=0.4\textwidth]{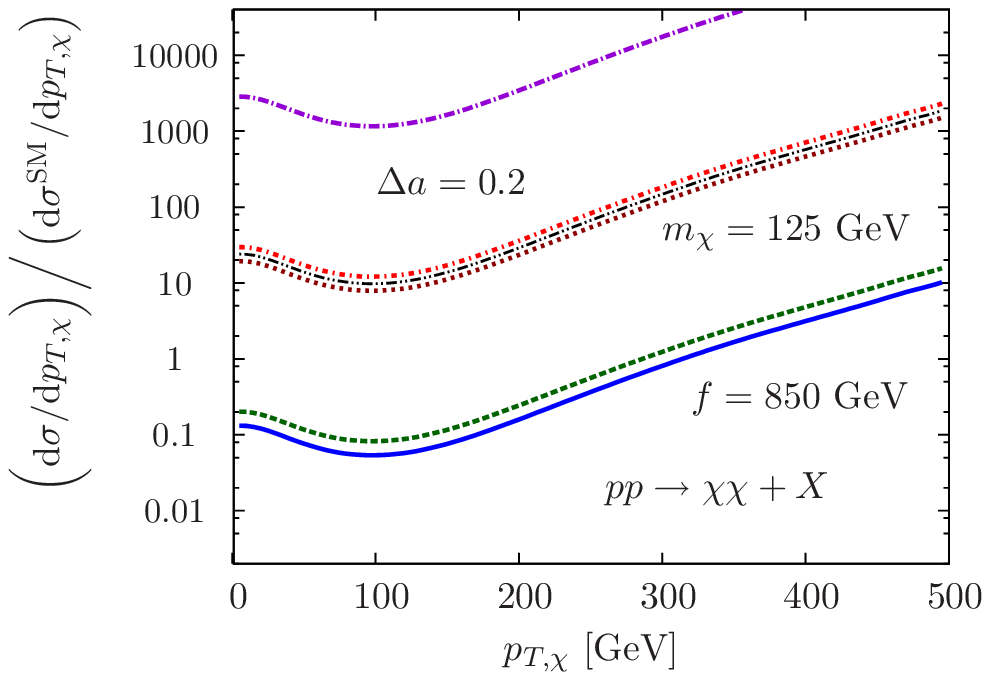} 
      \caption{\label{fig:pt_dilaton}
        Comparison of $\sigma(\chi\chi)\times {\text{BR}}(\chi_1)
        {\text{BR}} (\chi_2)$ to the values of the SM as a function of
        $p_{T,\chi}$ for $\Delta a=0.05$ (left panel) and $\Delta a=0.2$
        (right panel) and $c_{S}=7,c_{\chi\chi GG}=1$. The
        comparison of $\Delta a=0.05,0.2$ is depicted in
        Fig.~\ref{fig:pt_dilaton3}.}
    }\hfill
    \end{center}
\end{figure*}

\begin{figure}[!b]
  \includegraphics[width=0.4\textwidth]{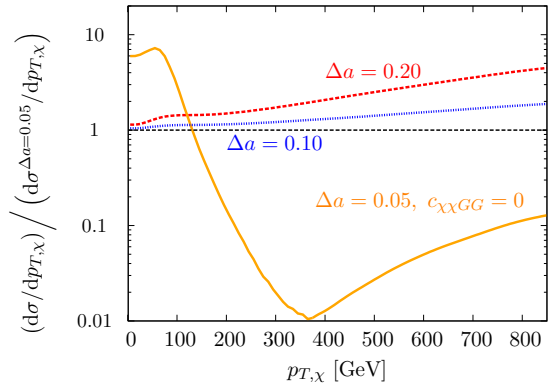}
  \caption{\label{fig:pt_dilaton3} Comparison of $\sigma(\chi\chi)$
    for different values of $\Delta a$ and $c_{\chi\chi GG}$ as a
    function of $p_{T,\chi}$ for $c_{S}=7$, $f=850~\gev$ fixed. The
    blue dotted line gives a comparison of $\Delta a=0.1$ to $\Delta
    a=0.05$ for $c_S,c_{\chi\chi GG}$ fixed, which and shows the
    dependence on the trilinear coupling.}
\end{figure}

We introduce explicit sources of scale symmetry
breaking~\citeb{Goldberger:2007zk} through the operator $
\lambda_{\mathcal{O}} \mathcal{O}(x)$, where the scaling dimension of
the operator $\mathcal{O} \neq 4$ induces non-derivative trilinear
interactions for $\chi$.
When the operator $\mathcal{O}$ is nearly marginal, so that its
anomalous dimension $\gamma = |\Delta_{\mathcal{O}} -4| \ll 1$ and one
writes the trilinear coupling as
$\frac{\lambda}{6}\frac{m_{\chi}^2}{f}\chi^3$, one obtains for
$\lambda$
\begin{equation}
  \lambda = (\Delta_{\mathcal{O}}+1) + \dots , \qquad \lambda_{\mathcal{O}} \ll 1
\end{equation}
where we must have at $\lambda \geq 2$ by the conformal algebra and
unitarity. If $\Delta_{\mathcal{O}}=2$ we obtain the Standard Model
result, rescaled by the ubiquitous factor of $v/f$.  Another
possibility is when $\gamma \ll 1$ when one obtains $\lambda =5$, 66\%
larger than the SM trilinear up to factors of $v/f$.  There are also
interesting anomalous four-derivative interactions in the low-energy
dilaton theory~\citeb{Komargodski:2011vj,Komargodski:2011xv},
\begin{equation}
  {\cal{L}}^{D7,D8} \supset 2(a_{UV}-a_{IR})  
  (2(\partial \chi)^2 \square \chi - (\partial \chi)^4),
  \label{eq:anomalous_derivative}
\end{equation}
of which the first gives rise to a trilinear interaction.  As these
interactions are derivative their largest effects will be seen in the
high $p_T$ regime, which we exploited in~\citeb{Dolan:2012rv} in order
to suppress backgrounds to a manageable level, If we consider a
strongly interacting $SU(N)$ gauge theory, then there will be $N^2-1$
gauge fields, and the theory will be approximately conformal if there
are $\sim 11N$ flavors of Weyl fermion. Taking $N=4,5,6$ we obtain
$a_{UV}= 0.033, 0.053$ and 0.076, using results
in~\citeb{Komargodski:2011vj}. We will initially take $\Delta a
=0.05$, but also consider a 'large' anomaly coefficient scenario,
where we take $\Delta a = 0.2$.

We summarize the parameter values we use regarding double dilaton
production in Table~\ref{tab:dilaton}, and show $c_{\chi\chi GG}$ in
brackets to indicate that we usually use the value derived by matching
with the effective field theory, but sometimes switch its effect off
altogether.

\begin{table}[h]
  \begin{center}
    \begin{tabular}{|c|c||c|c|} \hline
      Parameter & Value  & Parameter & Value \\ \hline
      $f$ & 850~GeV & $\lambda$ & 3 (SM)\\ \hline
      $c_{S}$ & 7 (4/3)& $c_{EM}$ & -17/9  (-1.2) \\ \hline
      $\Delta a$ & 0.05 (0.2) & $c_{\chi\chi GG}$ & (0) \\ \hline
    \end{tabular}
  \end{center}
  \caption{Parameters used in the calculation of double-dilaton 
    production in Section~\ref{sec:dilaton}\label{tab:dilaton}}
\end{table}

\begin{figure}[!b]
  \vspace{0.3cm}
  \includegraphics[width=0.21\textwidth]{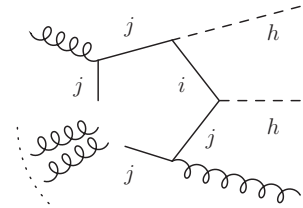}
  \caption{\label{fig:comgrp} Schematic representation of the $2h+n g$
    irreducible one-loop (sub)amplitude and for the involved fermion
    flavors in MCHM5, the gluon lines should be understood as
    off-shell currents contributing to {\it e.g.} $q\bar q \to hhg$. The
    amplitudes involving the trilinear Higgs vertex ({\it i.e.} the
    irreducible $h+ng$ (sub)amplitudes) are flavor diagonal due to
    diagonality at the gluon vertices $\slashed{A}\bar f_i f_j \propto
    \delta_{ij}$. We include all partonic subprocesses in our
    calculation.}
\end{figure}

Figure~\ref{fig:pt_dilaton} shows the differential distribution of
$\sigma \times {\text{BR}}$ for a number of final states, normalized
to those of the SM, in both the low and high anomaly coefficient
scenarios. The lower panels show the fully composite SM and the upper
ones the four family scenario. The effects of the higher dimensional
operators changing the $p_T$ spectrum can be seen entering at around
150~GeV.  In the fully composite case, while the cross-section for
those final states involving either 2 or 4 gluons are boosted with
respect to the SM, the final states that have proved useful in
previous dihiggs analyses are suppressed relative to the SM, even
though the total cross-section for $\chi\chi$ is considerably
higher. This is due to the double suppression coming from the factor
$v^2/f^2$ associated with massive final states. Although the
$\gamma\gamma jj$ final state cross-section is ten times the SM rate,
the leading order background is still too large to make an effective
analysis.  As it will never be feasible to pick out the relatively few
$gggg$ or $bbgg$ events from the enormous QCD background, one does not
expect any signal for this particular scenario.  One possible
exception is in the very boosted regime where $p_{T,\chi} \geq
350$~GeV, if the effects of higher dimensional operators are large.

On the other hand, the suppression factor into massive states is
smaller in the four family case, and the overall branching ratios are
more similar to their SM values. While the extra top-partners enhance
the total rate, the branching ratio to gluons is not so enhanced so as
to render an analysis impossible. On the contrary, $\sigma \times
{\text{BR}}$ for $bb\tau\tau$ and $bbW^+ W^-$ is approximately an
order of magnitude larger than in the SM, a factor which is enhanced
even more in the high $p_T$ tail of the distribution.

In Fig.~\ref{fig:pt_dilaton3} we show the effects on the $p_T$
differential distribution of varying the anomaly coefficient $\Delta
a$ and the dimension 6 coefficient $c_{\chi\chi GG}$, relative to the
'standard' case with $\Delta a =0.05$. The yellow line includes only
the anomalous derivative couplings which appeared in the proof of the
a-theorem. Its effect is similarly boosted in the low $p_T$ region
where there is lack of destructive interference due to the absence of
extra box-diagrams. We see that the effects of these interactions
becomes important for $p_{T,\chi} \sim 350$~GeV, where it can change
the cross-section by a factor of a few. The prospects for using the
di-dilaton final state to constrain the properties of the theory's UV
completion are thus promising.

\bigskip

{\it Summary:} The cross-section for di-dilaton production is much
larger than in the Standard Model. However, the future LHC prospects
for this scenario exhibit a strong dependence on ones assumptions
about the UV properties of the theory. In the fully composite SM, when
the suppression associated with non-gluonic final states is taken into
account, all possibly observable final states are too suppressed by
their branching ratios to give a signal at the LHC. On the other hand,
in the four-family scenario the prospects are excellent, with the
cross-section for reconstructible final states enhanced by up to an
order of magnitude. This is large enough that one may begin to
constrain further facets of the UV theory which manifest themselves
through higher dimensional operators.

\subsection{Composite di-Higgs production} 
The other possibility to have a light SM-like Higgs boson that we
discuss in this work is the composite Higgs scenario. The composite
Higgs~\citeb{Agashe:2004rs,Giudice:2007fh} relies on gauging the
electroweak interactions as a subgroup of a larger spontaneously
broken global symmetry group, {\it{e.g.}}
\begin{equation}
  \label{eq:symmb}
  SO(5)\to SO(4)\simeq SU(2)_L\times SU(2)_R\,, 
\end{equation}
which contains the gauged $SU(2)_L$. Gauging a subgroup is tantamount
to explicit breaking of the global symmetry, and the (uneaten)
Nambu-Goldstone bosons that arise from global symmetry breaking pick up
a mass from a Coleman-Weinberg potential~\citeb{cw} that involves both
gauge and fermion loops and breaks electroweak
symmetry~\citeb{Agashe:2005dk,Agashe:2004rs,Contino:2006qr}.

To incorporate proper hypercharges we need to extend the symmetry
group to $SO(5)\times U(1)_X$, and we identify hypercharge as
$Y=X+T^3_R$ like in other models of strong symmetry
breaking~\citeb{Csaki:2003zu}. This mechanism is elegantly described
by holographic approaches~\citeb{Contino:2003ve}, where symmetry
breaking is realized via the Hosotani
mechanism~\citeb{Hosotani:1983xw} in gauge-Higgs unified models.

%
%
\begin{figure}[!t]
  \includegraphics[height=0.4\textwidth]{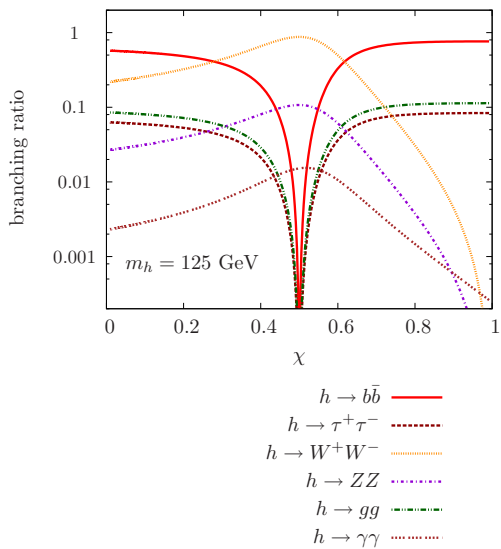}
  \caption{\label{fig:br} Branching ratios for the $m_h=125$~GeV
    Higgs as a function of $\chi$ in MCHM5.}
\end{figure}

The crucial parameter that measures deviations of the physical Higgs'
couplings to SM matter and parametrizes the model's oblique
corrections, is given by $\xi=v^2/f^2$, where $f$ is the analogue to
the pion decay constant. Consistency with experimental data can be
achieved without tuning, which makes this model class a promising
candidate for a BSM Higgs sector. In these composite Higgs models one
generates fermion masses (at least partially) via linear mixings with
composite fermionic operators instead of Technicolor-type interactions
to avoid bounds $\xi\ll 1$. In total, this amounts to a highly
modified di-Higgs phenomenology compared to the SM expectation, which
has already been discussed in
Refs.~\citeb{composite,composite2,Contino:2010mh,Espinosa:2010vn} in
some detail. In Ref.~\citeb{gillioz}, the effects of the light
additional fermionic degrees of freedom in the minimal composite Higgs
model based on Eq.~\gl{eq:symmb} (referred to as MCHM5) have been
included to inclusive di-Higgs predictions beyond LET (see also
Ref.~\citeb{dawson}). The additional fermions that run in the gluon
fusion loops strongly enhance the cross section, and, therefore, can
be highly constrained by applying the strategies that involve jet
recoils in di-Higgs production discussed in our previous
paper~\citeb{Dolan:2012rv} as we will see below.

MCHM5 introduces a set of composite vector-like fermions that form a
complete $5$ under $SO(5)$. The $5$ decomposes under the unbroken
$SU(2)_L\times SU(2)_R$, $\psi\equiv
5_{2/3}=(2,2)_{2/3}+(1,1)_{2/3}$. Obviously, the $5_{2/3}$ contains a
weak doublet of fields with the same quantum numbers as the
left-handed SM quark doublet $q_L=(t_L,b_L)^T$ and right-handed top
quark, and we can interpret the large mass of the top quark as a
mixing effect,
\begin{subequations}
  \label{eq:mass}
  \begin{multline}
    -{\cal{L}}_m =  y f (\bar \psi_L \Sigma^T)(\Sigma \psi_R) + m_0
    \bar \psi_L \psi_R \\  + \Delta_L \bar q_L Q_R +\Delta_R 
    \overline{\widetilde{T}}_L t_R + {\text{h.c.}}\,,
  \end{multline}
  where the non-linear Higgs field $\Sigma$ is parametrized via the
  $SO(5)/SO(4)$ coset space generators and can be chosen (see
  {\it{e.g.}} Ref.~\citeb{gillioz})
  \begin{equation}
    \Sigma=(0,0,\sin(h/f),0,\cos(h/f))\,.
  \end{equation}
\end{subequations}
Expanding the non-linear sigma model we recover the interactions with
electroweak gauge bosons as well as the Higgs self-couplings relevant
to this study
\begin{multline}
  {\cal{L}}_{h} = {1\over 2} (\partial_\mu h)^2 -
  {m_h^2\over 2}\,h^2 - {1-2\xi\over \sqrt{1-\xi}}h^3 + \dots \\
 + {g^2f^2\over 4}\sin^2\left({h\over f}\right) \left( W^+_\mu W^{-\,\mu}
    + {1\over \cos^2\theta_w} Z_\mu Z^\mu \right)\,,
\end{multline} {\it{i.e.}} we need to rescale the SM trilinear $hVV$
vertices by a factor $\sqrt{1-\xi}$ and we have $ f^2 \sin^2(\left\langle
  h\right\rangle /f)=v$. The Higgs branching ratios of MCHM5 are
depicted in Fig.~\ref{fig:br}.

Following Ref.~\citeb{gillioz}, we do not include another $5_{-1/3}$
multiplet for generating the bottom quark mass, but include it by
breaking partial compositeness with an explicit coupling of the
Yukawa-like interactions. Expanding Eq.~\gl{eq:mass} in the mass
diagonal basis, we obtain the masses of the fermionic mass spectrum and
interactions $h\bar f_i f_j$ and $hh\bar f_if_j$ (where $i,j$ run over
the heavy fermion flavors) which are relevant for di-Higgs(+jet) production
from gluon fusion, which is the dominant production
mechanism\footnote{Di-Higgs production from weak boson
  fusion~\cite{wbfdi} is suppressed, also because in addition to the
  $hVV$ vertices the $hhVV$ vertices are rescaled by $1-2\xi$ with
  respect to the SM. Unitarization of the $V_LV_L\to V_LV_L,q\bar q$
  amplitudes is partially taken over by the exchange of techni-$\rho$
  like resonances. These can be studied in the weak boson fusion
  channels~\cite{Bagger:1993zf}.}.

In general, the composite Higgs interactions Eq.~\gl{eq:mass} will not
be flavor-diagonal in the space of states that contains the composite
multiplet augmented by $t_{L,R}$, and constraints from both direct
detection of flavor measurements are eminent. For the remainder of
this section we will choose parameter points that are in agreement
with these constraints to discuss the composite Higgs model's
implications on di-Higgs and di-Higgs+jet phenomenology following
Ref.~\citeb{gillioz}.

We take into account all non-diagonal couplings and keep the
full mass dependence in the calculation beyond any approximation. This
results in computationally intense calculations, especially for the
pentagon part in $gg\to g hh$ and box $gg \to hh$ (sub)amplitudes
where non-diagonality of the $h\bar f_if_j$ vertices increases the
feynman graph combinatorics, Fig.~\ref{fig:comgrp}.

\begin{figure}[!t]
  \includegraphics[width=0.4\textwidth]{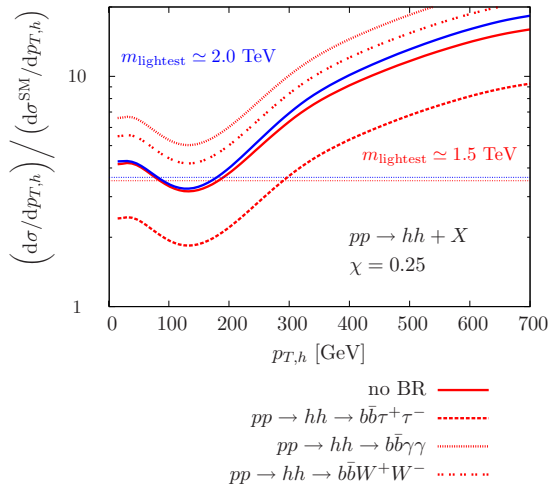}
  \caption{\label{fig:pt_comp} Comparison of composite di-Higgs
    production $p_{T,H}$ spectra with the SM for $\xi=0.25$.}
\end{figure}

The result in comparison to the SM is shown in Fig.~\ref{fig:pt_comp}
for $pp\to hh+X$ production. For a typical mass spectrum $m_t\simeq
174~\gev$ and the lightest composite fermion
$m_{\text{lightest}}\simeq 1500~\gev$ we find agreement with the
enhanced cross sections as reported in Ref.~\citeb{gillioz},
$\sigma(hh)/\sigma^{\text{SM}}(hh)\sim 3$. The phase space dependence
of this enhancement is rich and non-trivial as a consequence of the
non-diagonal couplings and additional mass scales that show up in the
box contributions, which also interfere with modified trilinear
interactions. Hence, it is difficult to comment on quantitative
similarities of the composite Higgs phenomenology for different
parameter choices. 

However, on a qualitative level, since the composite scale needs
typically to be large in order to have agreement with direct searches
and flavor bounds, the inclusive $pp\to hh+X$ composite phenomenology
will be dominated by modifications with respect to the SM at medium
$p_{T,h}\simeq 100~\gev$. This phase space region is mostly sensitive
to modifications of the $tth$ coupling and the modified trilinear $h$
vertex. At large $p_{T,h}$ we observe an enhancement due to the
presence of new massive fermions in the box contributions of the
$\stackrel{\text{\tiny(}-\text{\tiny)}}{q} \stackrel{}{g}$-initiated
subprocesses, which access the protons' valence quark distribution. We
note that higher order QCD corrections are likely to further enhance
the cross section prediction beyond the naive
SM-rescaling~\citeb{hpair,Furlan:2011uq}.

We find an even larger enhancement of leading order $pp\to
hh+{\text{jet}}$ production cross section, with $p_{T,j}\geq 80~\gev$
\begin{equation}
  \sigma(hh+j) \simeq 13.0~\text{fb}\,,
\end{equation}
for both scenarios shown in Fig.~\ref{fig:pt_comp}. This result needs
to be compared to the corresponding LO prediction in the SM which is
$\sigma^{\text{SM}}=2.8~\text{fb}$, and amounts to an enhancement of a
factor of $4.6$. For the fully hadronized search of
Ref.~\cite{Dolan:2012rv} this amounts to $S/B\simeq 7$, which is well
beyond systematic background uncertainties for high luminosity
searches.

The relatively larger increase of the one jet-inclusive cross section
can be understood along the following lines. The additional top
partners introduce a new mass scale to the one-loop amplitude. At
large transverse momentum, the cross section is dominated by continuum
$hh$ production which mostly proceeds via box diagrams in addition to
initial radiation. The latter is increased as a result of the newly
introduced mass scale in comparison to the SM, and initial state
radiation allows the initial state partons to access the large valence
quark parton distributions. This effect is also visible in the NLO
predictions of $pp\to hh+X$ in composite models employing the
effective theory approximation~\cite{Furlan:2011uq}.

\bigskip
{\it Summary:} 

The composite Higgs scenario is a well-motivated model of electroweak
symmetry breaking that is consistent with current flavor constraints
and direct searches for heavy top partners. Furthermore, composite
Higgs models typically predict a large enhanced di-Higgs cross
section, which is further enhanced in for $hh+{\text{jet}}$ final
state by the introducing a new mass scale to the phenomenology. While
small di-Higgs(+jet) rates in the context of the SM might hinder a
determination of the SM Higgs potential in case no further indications
of physics beyond the SM become available, composite di-Higgs
production will overcome this shortcoming due to its large production
cross section. Consequently, also for extremely heavy top partners,
di-Higgs(+jet) production is going to provide a powerful test of Higgs
compositeness at the LHC.

\section{Conclusions}
A precise determination of the realization of Higgs mechanism {\it sui
  generis} is an important task that has to be pursued at the LHC,
especially after the recent discovery of an SM Higgs-like
particle. While measurements based on single Higgs boson production
provide only indirect constraints on the realization of electroweak
symmetry breaking, the partial experimental reconstruction of the
Higgs potential is indispensable to gain a fuller understanding at a more
fundamental level.

In this paper, we have investigated di-Higgs and di-Higgs+jet production
in a variety of model classes, whose single Higgs production
characteristics can account for the observation of the new particle at
the LHC. Rather than employing an agnostic field theory
approach\footnote{See Ref.~\citeb{dihiggs2} for related discussions.}
we have picked well-motivated examples of realistic BSM (scalar)
sectors, supplemented by the required fermionic particle content,
which generalize the SM Higgs sector in two fundamentally distinct
ways.

The first option deals with models with extended Higgs sectors
predicting new resonant structures in di-Higgs production due to the
model's two-Higgs doublet character. Furthermore, new kinematical
configurations can provide extra analytical handles in the production
of a heavy Higgs boson in addition to the light Higgs. In
portal-inspired scenarios, the determination of the involved trilinear
couplings is important to reconstruct the full extended portal
potential for parameter points where the two Higgs bosons are not too
widely separated in mass. In the MSSM, a corresponding measurement
facilitates the reconstruction of the Higgs sector mixing angles
$\alpha$ and $\beta$, and hence provides indirect constraints on the
stop masses and mixing parameter $A_t$. This can be achieved by
separating the resonant contribution from continuum production via
invariant mass cuts, and applying boosted~\cite{Dolan:2012rv} and
unboosted~\cite{Baur:2003gp} analysis strategies to the different
samples.

\bigskip

The resonant models are contrasted to realizations of the Higgs
mechanism where the ``Higgs'' boson arises as a pseudo Nambu-Goldstone
mode of some spontaneously broken symmetry. The agreement of current
observations with the SM Higgs predictions requires the pseudo-Nambu-Goldstone
boson to have similar couplings as the SM Higgs boson. Along with
composite Higgs models this leaves only the pseudo-dilaton as a second
option.

The former case implies interpreting the entire Higgs doublet as a set
of Nambu-Goldstone fields. Realistic composite Higgs scenarios predict
strongly-enhanced di-Higgs and di-Higgs+jet cross sections. 

In models with an approximate conformal invariance, symmetry breaking
can be triggered at scales considerably higher and spontaneous
breaking of conformal invariance introduces a new light state to the
low energy effective theory, which has similar properties as the SM
Higgs boson as a consequence of Eq.~\gl{eq:coupling}: the
pseudo-dilaton. For both the composite and the dilaton option, there
are parameter choices such that current observations can be accounted
for.  It is their highly modified di-Higgs phenomenology which can
effectively discriminates these possibilities depending on the further
particularities of the conformal sector, and facilitates an LHC
measurement of the involved couplings and parameters in case of the
composite Higgs model. Pseudo--di-dilaton production can be buried in
a large hadronic background with no kinematic handles to reconstruct
the preferred dilaton decay to gluons. In this sense, the absence of a
``traditional'' di-Higgs phenomenology could be interpreted as
evidence for a dilatonic realization. Interpreting the presence of a
large di-Higgs(+jet) production cross section is more involved, and
could be evidence for a fourth-family (or more complicated)
realization of the pseudo-dilaton model, but may also be consistent
with a composite Higgs.

It is clear from our analysis that, no matter what governs the
dynamics of the newly-discovered boson, its multi-production
phenomenology, which can be studied at the LHC in sufficient detail,
will provide a clear image of its role in the mechanism of electroweak
symmetry breaking. These findings will further consolidate with an LHC
luminosity upgrade \cite{atlashh}.

\bigskip 
{\bf{Acknowledgments}} ---
We thank John Ellis, Christophe Grojean and Peter M. Zerwas for helpful
discussions.
CE acknowledges funding by the Durham International Junior Research
Fellowship scheme.


\end{document}